\documentclass[10pt,a4paper,conference]{IEEEtran}
\usepackage[dvipdfm]{graphicx}
\usepackage{bm}
\usepackage{amsmath}
\usepackage{amssymb}

\makeatletter
\def\cosec{\mathop{\operator@font cosec}\nolimits}
\def\sech{\mathop{\operator@font sech}\nolimits}
\def\cosech{\mathop{\operator@font cosech}\nolimits}
\def\atanh{\mathop{\operator@font atanh}\nolimits}
\def\arctanh{\mathop{\operator@font arctanh}\nolimits}
\def\arcsec{\mathop{\operator@font arcsec}\nolimits}
\def\arccosec{\mathop{\operator@font arccosec}\nolimits}
\def\arccosh{\mathop{\operator@font arccosh}\nolimits}
\def\arcsinh{\mathop{\operator@font arcsinh}\nolimits}
\def\arcsech{\mathop{\operator@font arcsech}\nolimits}
\def\arccosech{\mathop{\operator@font arccosech}\nolimits}
\def\E{\mathop{\operator@font E}\nolimits}
\def\I{\mathop{\operator@font I}\nolimits}
\def\KL{\mathop{\operator@font KL}\nolimits}
\def\tr{\mathop{\operator@font tr}\nolimits}
\def\Var{\mathop{\operator@font Var}\nolimits}
\def\erf{\mathop{\operator@font erf}\nolimits}
\def\ext{\mathop{\operator@font ext}}
\def\extr{\mathop{\operator@font extr}}
\def\sign{\mathop{\operator@font sign}\nolimits}
\def\const{\mathop{\operator@font const}\nolimits}
\makeatother
\newcommand\Eq[1]{(\ref{eq:#1})}
\def\EbN0{E_\mathrm{b}/N_0}
\begin{document}

\title{An LDPCC decoding algorithm\\ based on Bowman-Levin approximation\\ ---Comparison with BP and CCCP---}

\author{
\authorblockN{Masato Inoue, Miho Komiya and Yoshiyuki Kabashima}
\authorblockA{
    Department of Computational Intelligence and Systems Science,\\
    Interdisciplinary Graduate School of Science and Engineering,\\
    Tokyo Institute of Technology, Yokohama 226-8502, Japan\\
    Email: inoue@sp.dis.titech.ac.jp, miho@sp.dis.titech.ac.jp, and kaba@dis.titech.ac.jp
}}

\maketitle

\begin{abstract}
Belief propagation (BP) and the concave convex procedure (CCCP) are both methods
that utilize the Bethe free energy as a cost function and solve information processing tasks.
We have developed a new algorithm that also uses the Bethe free energy,
but changes the roles of the master variables and the slave variables.
This is called the Bowman-Levin (BL) approximation in the domain of statistical physics.
When we applied the BL algorithm to decode the Gallager ensemble of short-length regular low-density parity check codes (LDPCC)
over an additive white Gaussian noise (AWGN) channel,
its average performance was somewhat better than that of either BP or CCCP.
This implies that the BL algorithm can also be successfully applied to 
other problems to which BP or CCCP has already been applied.
\end{abstract}

%================================================================
\section{Introduction}
  Recently, various statistical inference algorithms have become of interest
in the field of large-scale information processing.
Belief propagation (BP) \cite{rf:Pearl} and the concave convex procedure (CCCP) \cite{rf:Yuille}
are among the most effective of the methods which minimize the Bethe free energy \cite{rf:Kabashima,rf:Yedidia}.
In the field of practical application
(e.g., the problem of decoding low-density parity check code (LDPCC) \cite{rf:Gallager,rf:MacKay}),
BP and CCCP have both been successfully applied \cite{rf:Shibuya}.

  However, they are not the only methods that minimize the Bethe free energy.
In this paper, we focus on the method of Lagrange undetermined multipliers used by both BP and CCCP,
and derive a new algorithm by exchanging the roles of master variables and slave variables.
This approach, called Bowman-Levin (BL) approximation \cite{rf:Bowman},
is sometimes used in the field of statistical physics
as a way to find an extremum (a saddle, local minimum, or local maximum) of the Bethe free energy.

%================================================================
\section{Low density parity check code (LDPCC)}
  The LDPCC decoding problem can be handled within a Bayesian framework.
The prior probability of the codes, consisting of $N$ binary bits (${\bm x}\in\{+1,-1\}^N$), is defined as
\begin{eqnarray}
    P({\bm x})     \propto \prod_\mu^M \left( 1\!+\!\prod_{l\in{\bm \mu}}x_l \right) ,
\end{eqnarray}
where $\mu=1,...,M$ denotes the parity index and ${\bm \mu}$ denotes the set of node indices involved in the $\mu$-th parity.
Similarly, $l=1,...,N$ denotes the bit index and ${\bm l}$ denotes the set of parity indices linking to the $l$-th bit.
$|{\bm \mu}|$ and $|{\bm l}|$ denote the degree of $\mu$-th parity and the $l$-th bit, respectively.
The proportion means the normalization of a probability function -- i.e., the summation of the probability for all possible arguments ${\bm x}$ -- should be $1$.

  We consider a noisy channel with additive white Gaussian noise (AWGN);
i.e., the probability of the received codes ${\bm y}$ is defined as
\begin{eqnarray}
    P({\bm y}|{\bm x}) \propto \prod_l^N \exp\left( -\frac{(y_l\!-\! x_l)^2}{2\sigma^2} \right) ,
\end{eqnarray}
where $\sigma^2$ denotes the variance of the noise.
The posterior probability of the sent code can then be expressed as
\begin{eqnarray}
    P({\bm x}|{\bm y})
    \propto \left[ \prod_\mu^M \left( 1\!+\!\prod_{l\in{\bm \mu}}x_l \right)  \right]\left[ \prod_l^N \exp\left( x_l\frac{y_l}{\sigma^2} \right) \right] .
\end{eqnarray}

  To infer the sent code ${\bm x}$ by ${\bm y}$,
we employ the maximum posterior marginal (MPM) solution,
\begin{eqnarray}
    {\hat{x}}_l = \arg\max_{x_l} \sum_{{\bm x}_{\backslash l}} P({\bm x}|{\bm y}) ,
\end{eqnarray}
which minimizes the bit error rate.
On the other hand, the maximum a posteriori (MAP) solution minimizes the block error rate,
\begin{eqnarray}
    {\hat{\bm x}} = \arg\max_{{\bm x}} P({\bm x}|{\bm y}) ,
\end{eqnarray}
but is generally difficult to determine because of the exponential calculation cost.

%================================================================
\section{Bethe free energy}
%================
\begin{figure}[t]
\begin{picture}(120,45)
\put( 20,32){(a)}
\put( 20, 0){\includegraphics[width=40mm]{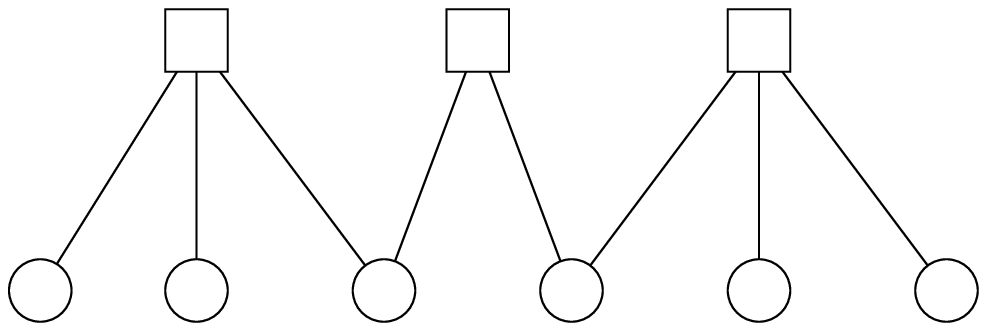}}
\put(140,30){parity checks}
\put(140, 2){bits}
\end{picture}\\%
\begin{picture}(120,45)
\put( 20,32){(b)}
\put( 38, 0){\includegraphics[width=31mm]{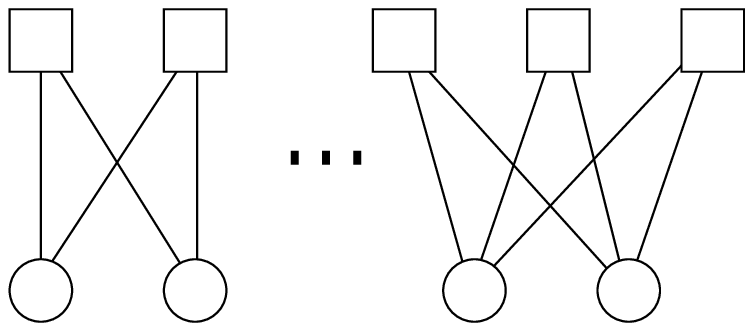}}
\put(140,30){parity checks}
\put(140, 2){bits}
\end{picture}%
    \caption{Examples of the parity connection. (a) Tree structure and (b) loopy structure.}
    \label{fig:Tree}
\end{figure}
%================
  One purpose of the Bethe free energy approach is 
to determine a set of marginal probabilities of 
a given probability, which provides the MPM solution here.
The Bethe free energy, $F$, is defined using Kullback-Leibler (KL) divergence as
\begin{eqnarray}
  \label{eq:BetheKL}
    F
    \!\!&\!\!\equiv\!\!&\!\! \sum_\mu^M \KL(b_\mu({\bm x}_{\bm \mu})||\phi_\mu({\bm x}_{\bm \mu}))  \\
    \!\!&\!\!      \!\!&\!\! + \sum_l^N (1\!-\!|{\bm l}|) \KL(q_l(x_l)||\psi_l(x_l)) , \nonumber
\end{eqnarray}
where $P({\bm x}|{\bm y})$ can be represented as
\begin{eqnarray}
    P({\bm x}|{\bm y})      \!\!&\!\!\propto\!\!&\!\! \left[ \prod_\mu^M \phi_\mu({\bm x}_{\bm \mu}) \right]\left[ \prod_l^N \psi_l(x_l)^{1\!-\!|{\bm l}|} \right], \\
    \phi_\mu({\bm x}_{\bm \mu}) \!\!&\!\!\propto\!\!&\!\! \left[ 1\!+\!\prod_{l\in{\bm \mu}}x_l \right]\left[\prod_l^N \psi_l(x_l) \right] ,\\
    \psi_l(x_l)        \!\!&\!\!\propto\!\!&\!\! \exp\left( x_l\frac{y_l}{\sigma^2} \right), 
\end{eqnarray}
using (normalized) probability functions, $\phi_\mu({\bm x}_{\bm \mu})$ and $\psi_l(x_l)$ in the current case.
The introduced test probability functions of creeks, $\{b_\mu({\bm x}_{\bm \mu})\}$, and bits, $\{q_l(x_l)\}$,
are required to satisfy the consistency of the marginal probabilities:
\begin{eqnarray}
    \sum_{{\bm x}_{{\bm \mu}\backslash l}} b_\mu({\bm x}_{\bm \mu}) = q_l(x_l) ~~~~ (l\in {\bm \mu}).
  \label{eq:consistency}
\end{eqnarray}
$\{b_\mu({\bm x}_{\bm \mu})\}$ and $\{q_l(x_l)\}$ that minimize $F$ are expected to approximate the marginal probabilities of $P({\bm x}|{\bm y})$.

  The Bethe free energy approach gives the exact marginal probabilities
when the parity connection has the tree structure (Fig. \ref{fig:Tree}(a)).
In such cases, any probability of ${\bm x}$ can be expressed as a product of 
its marginal probabilities, $\{b_\mu({\bm x}_{\bm \mu})\}$ and $\{q_l(x_l)\}$, as
\begin{eqnarray}
    Q({\bm x})      \!\!&\!\!   =  \!\!&\!\! \left[ \prod_\mu^M b_\mu({\bm x}_{\bm \mu}) \right]\left[ \prod_l^N q_l(x_l)^{1\!-\!|{\bm l}|} \right] .
\end{eqnarray}
Then, the Bethe free energy coincides with the KL divergence between 
the test and the true probabilities,
\begin{eqnarray}
    F = \KL(Q({\bm x})||P({\bm x}|{\bm y})), 
\end{eqnarray}
which implies that minimizing the Bethe free energy leads to the correct probability $Q({\bm x})=P({\bm x}|{\bm y})$,
and, therefore, the exact MPM solution can be assessed 
from the obtained $\{q_l(x_l)\}$.
Unfortunately, the Bethe free energy does not represent the KL divergence for loopy graphs (Fig. \ref{fig:Tree}(b)).
However, we here attempt to decode the LDPCC by minimizing $F$ under the consistency condition \Eq{consistency}
expecting that $\{q_l(x_l)\}$ well approximate the marginal probabilities
even in the case that the parity connection does not have the tree structure.

%================================================================
\section{Lagrange multipliers}
  To minimize the Bethe free energy
under the constraint \Eq{consistency}, we introduce 
Lagrange undetermined multipliers, $\lambda_{\mu l}(x_l)$.
The objective function to minimize is
\begin{eqnarray}
    G(\{b_\mu\},\{q_l\},\{\lambda_{\mu l}\})
    \equiv F + L ,
\end{eqnarray}
where
\begin{eqnarray}
    L \equiv 
    \sum_\mu^M \sum_{l\in {\bm \mu}} \sum_{x_l} \lambda_{\mu l}(x_l) \left( \sum_{{\bm x}_{{\bm \mu}\backslash l}} b_\mu({\bm x}_{\bm \mu}) - q_l(x_l) \right) ,
\end{eqnarray}
and we solve the following three equations:
\begin{eqnarray}
  \label{eq:ob}
    0 \!\!&\!\!   =  \!\!&\!\!
    \frac{\partial G}{\partial b_\mu({\bm x}_{\bm \mu})} = \ln\frac{b_\mu({\bm x}_{\bm \mu})}{\phi_\mu({\bm x}_{\bm \mu})} + 1 + \sum_{l\in {\bm \mu}} \lambda_{\mu l}(x_l) , \\
  \label{eq:oq}
    0 \!\!&\!\!   =  \!\!&\!\!
    \frac{\partial G}{\partial q_l(x_l)} = (1\!-\!|{\bm l}|)\left( \ln\frac{q_l(x_l)}{\psi_l(x_l)} + 1  \right) - \sum_{\mu\in {\bm l}} \lambda_{\mu l}(x_l) , \nonumber\\ \\
  \label{eq:ol}
    0 \!\!&\!\!   =  \!\!&\!\!
    \frac{\partial G}{\partial \lambda_{\mu l}(x_l)} = \sum_{{\bm x}_{{\bm \mu}\backslash l}} b_\mu({\bm x}_{\bm \mu}) - q_l(x_l) .
\end{eqnarray}

  Using $x_l\in\{+1,-1\}$ and the normalization conditions of the probability functions,
we can reduce $q_l$ and $\lambda_{\mu l}$ to linear functions as
\begin{eqnarray}
    q_l(x_l) \!\!&\!\!   =  \!\!&\!\! \frac{1\!+\! x_l \tanh h_l}{2} ,\\
  \label{eq:lh}
    \lambda_{\mu l}(x_l) \!\!&\!\!   =  \!\!&\!\! -x_l \left( h_{\mu l} \!-\! \frac{y_l}{\sigma^2} \right) + \eta_{\mu l} .
\end{eqnarray}
We also sometimes use $m_l \equiv \left< x_l \right>_{q_l(x_l)} = \tanh h_l$.
Using these expressions, we can reduce Eqs. \Eq{ob} - \Eq{ol} to
\begin{eqnarray}
  \label{eq:b}
    b_\mu({\bm x}_{\bm \mu}) \!\!&\!\!\propto\!\!&\!\! \left[ 1\!+\!\prod_{l\in{\bm \mu}}x_l \right]\left[\prod_l^N \exp(x_l h_{\mu l}) \right],\\
  \label{eq:q}
    h_l \!\!&\!\!   =  \!\!&\!\!
    \frac{1}{|{\bm l}|\!-\!1} \left( \sum_{\mu'\in {\bm l}} h_{\mu' l} - \frac{y_l}{\sigma^2} \right) , \\
  \label{eq:l}
    m_l \!\!&\!\!   =  \!\!&\!\!
    \frac{\sum_{x_l} x_l \sum_{{\bm x}_{{\bm \mu}\backslash l}} b_\mu({\bm x}_{\bm \mu})}
       {\sum_{x_l}     \sum_{{\bm x}_{{\bm \mu}\backslash l}} b_\mu({\bm x}_{\bm \mu})} ,
\end{eqnarray}
respectively. From Eqs. \Eq{b} and \Eq{l}, we obtain
\begin{eqnarray}
  \label{eq:bl}
    h_l \!\!&\!\!   =  \!\!&\!\!
    h_{\mu l} + \atanh \prod_{l'\in {\bm \mu} \backslash l} \tanh h_{\mu l'}
\end{eqnarray}
Now, we have two types of variable: $\{h_l\}$ and $\{h_{\mu l}\}$,
and two types of simultaneous equation: \Eq{q} and \Eq{bl}.

%================================================================
\section{Belief propagation (BP)}
  BP considers the double-indexed $h$, $\{h_{\mu l}\}$, to be the master variables.
Specifically, from Eqs. \Eq{q} and \Eq{bl}, we obtain 
\begin{eqnarray}
    &&\hspace{15mm} h_{\mu'l} + \atanh \prod_{l'\in {\bm \mu}'\backslash l} \tanh h_{\mu'l'} \hspace{2mm} \nonumber\\
    &&\hspace{25mm} =
    \frac{1}{|{\bm l}|\!-\!1} \left( \sum_{\mu''\in {\bm l}} h_{\mu'' l} - \frac{y_l}{\sigma^2} \right)
\end{eqnarray}
for any $\{l, \mu'\in {\bm l}\}$.
BP ingeniously rearranges the left side of this equation with the average without $\mu$:
\begin{eqnarray}
    &&\hspace{-7mm} \frac{1}{|{\bm l}|\!-\!1} \sum_{\mu'\in {\bm l}\backslash\mu} \left( h_{\mu' l} + \atanh \prod_{l'\in {\bm \mu}' \backslash l} \tanh h_{\mu' l'} \right) \nonumber\\
    &&\hspace{25mm} =
    \frac{1}{|{\bm l}|\!-\!1} \left( \sum_{\mu''\in {\bm l}} h_{\mu'' l} - \frac{y_l}{\sigma^2} \right)
\end{eqnarray}
We then obtain the iterative substitution to converge $\{h_{\mu l}\}$.
\begin{eqnarray}
  \label{eq:BP}
    \textrm{loop: } h_{\mu l} \gets \frac{y_l}{\sigma^2} + \sum_{\mu'\in {\bm l}\backslash\mu} \atanh \prod_{l'\in {\bm \mu}'\backslash l} \tanh h_{\mu'l'}.
\end{eqnarray}
Once the master variables are determined, we can easily obtain the slave variables, $\{h_l\}$, by
\begin{eqnarray}
    \textrm{result: } h_l       \gets \frac{y_l}{\sigma^2} + \sum_{\mu'\in {\bm l}}       \atanh \prod_{l'\in {\bm \mu}'\backslash l} \tanh h_{\mu'l'}.
\end{eqnarray}

  To lower the calculation cost,
we check whether the estimated sent code, ${\hat{x}}_l \equiv \sign h_l$, satisfies all parities for every loop of Eq. \Eq{BP}.
We stop the iteration loop if we reach any codeword, or the number of loops reaches an upper limit.

%================================================================
\section{Concave convex procedure (CCCP)}
  CCCP is a double loop algorithm utilizing convex optimization.
The convexity of the Bethe free energy is generally not guaranteed because of the negative coefficient, $1\!-\!|{\bm l}|$, of the second term in Eq. \Eq{BetheKL}.
So, CCCP employs the following additional term at every outer loop step $t$.
\begin{eqnarray}
  \label{eq:CCCP}
    {\tilde{F}}^t
    \!\!&\!\!\equiv\!\!&\!\! F + \sum_l^N |{\bm l}| \KL(q_l(x_l)||q_l^t(x_l)) \\
  \label{eq:CCCPX}
    \!\!&\!\!   =  \!\!&\!\! \sum_\mu^M \KL(b_\mu({\bm x}_{\bm \mu})||\phi_\mu({\bm x}_{\bm \mu})) + \sum_l^N \KL(q_l(x_l)||\psi_l(x_l)) \nonumber\\
    \!\!&\!\!      \!\!&\!\! + \sum_l^N |{\bm l}| q_l(x_l) \ln\frac{\psi_l(x_l)}{q_l^t(x_l)} .
\end{eqnarray}
Equation \Eq{CCCPX} guarantees the convexity of ${\tilde{F}}^t(\{b_\mu\},\{q_l\})$,
because $\KL$ divergence function is convex, and the third term is a linear function.
Besides, $F$ necessarily decreases if ${\tilde{F}}^t$ decreases because the additional term is non-negative,
and the additional term itself disappears if $\{q_l\}$ converges.

  In the inner loop, similar to BP, CCCP considers the double-indexed $h$, $\{h_{\mu l}\}$, to be the master variables.
On the other hand, in the outer loop, single-indexed $h$, $\{h_l\}$, are treated as the master variables.
After the convergence of the inner loop, the outer loop is performed to determine $h_l^{t+1}$. 
\begin{eqnarray}
    &&\hspace{-5mm} \textrm{inner loop: } h_{\mu l} \gets \frac{1}{2} \left( \frac{y_l}{\sigma^2} + \sum_{\mu'\in {\bm l}\backslash\mu} (h_l^t-h_{\mu' l})  \right. \nonumber\\
    &&\hspace{25mm}                                                 \left. + h_l^t - \atanh \prod_{l'\in {\bm \mu} \backslash l} \tanh h_{\mu l'} \right)\!, \\
    &&\hspace{-5mm} \textrm{outer loop: } h_l^{t+1}   \gets \hspace{4mm}  \frac{y_l}{\sigma^2} + \sum_{\mu'\in {\bm l}      } (h_l^t-h_{\mu' l}) .
\end{eqnarray}

%================================================================
\section{Bowman-Levin (BL)}
  BL considers the single-indexed $h$, $\{h_l\}$, to be the master variables.
Specifically, BL determines $\{h_{\mu l}\}$ by first using $\{h_l\}$ in Eq. \Eq{bl}.
It requires some iteration to be solved, resulting in an inner loop:
\begin{eqnarray}
    \textrm{inner loop: } h_{\mu l} \gets h_l^t - \atanh \prod_{l'\in {\bm \mu} \backslash l} \tanh h_{\mu l'} .
\end{eqnarray}
Because the determination of the slave variables, $\{h_{\mu l}\}$, depends on the provisional values of the master variables, 
$\{h_l\}$,
BL also needs a double-loop algorithm.
Eq. \Eq{q} implies that update 
\begin{eqnarray}
    \textrm{outer loop: } h_l^{t+1} \gets \frac{1}{|{\bm l}|\!-\!1} \left( \sum_{\mu'\in {\bm l}} h_{\mu' l} - \frac{y_l}{\sigma^2} \right), 
  \label{eq:natural_OL}
\end{eqnarray}
may be employed for the outer-loop using the converged variables $\{h_{\mu l}\}$.

Eq. \Eq{natural_OL}, however, does not provides satisfactory results
as this empirically increases the Bethe free energy.
This is because the outer loop \Eq{natural_OL} is interpreted as
\begin{eqnarray}
    h_l^{t+1} \gets h_l^t +  \kappa \frac{\partial G^t}{\partial h_l} .
\end{eqnarray}
where $\kappa \equiv 
\frac{\cosh^{-2} h_l^t}{|{\bm l}|\!-\!1}$ is positive.
$G^t$ denotes $G$, which is regarded as a function of only $\{h_l\}$ at outer-loop step $t$.

In order to resolve this difficulty, we use 
the natural gradient descent method \cite{rf:Amari}
instead of Eq. \Eq{natural_OL} as
\begin{eqnarray}
    && \textrm{outer loop: } {\bm h}^{t+1} \gets {\bm h}_l^t - k {\bm H}^{-1} \frac{\partial G^t}{\partial {\bm h}},
\end{eqnarray}
where $k$ denotes a small positive step width,
and ${\bm H}$ denotes the Fisher information matrix defined as
\begin{eqnarray}
    H_{i,j}
    \!\!&\!\!\equiv\!\!&\!\! \left< \frac{\partial \log Q({\bm x})}{\partial h_i} \frac{\partial \log Q({\bm x})}{\partial h_j} \right>_{Q({\bm x})} \\
    \!\!&\!\!   =  \!\!&\!\! \left\{ \begin{array}{ll} \cosh^{-2} h_i & (i=j) \\ 0 & (i\neq j) \end{array} \right.
\end{eqnarray}
assuming the following approximation:
\begin{eqnarray}
    Q({\bm x}) \simeq \prod_l^N q_l(x_l) .
\end{eqnarray}
We then obtain
\begin{eqnarray}
    \textrm{outer loop: } h_l^{t+1} \gets h_l^t - k \frac{\partial G^t}{\partial m_l},
\end{eqnarray}
where
\begin{eqnarray}
    \frac{\partial G^t}{\partial m_l} = \sum_{\mu\in {\bm l}} h_{\mu l} - \frac{y_l}{\sigma^2} - (|{\bm l}|\!-\!1) h_l^t .
\end{eqnarray}

%================================================================
\section{Validation}

%================
\begin{figure}[t]
    \centering
    \includegraphics[width=80mm]{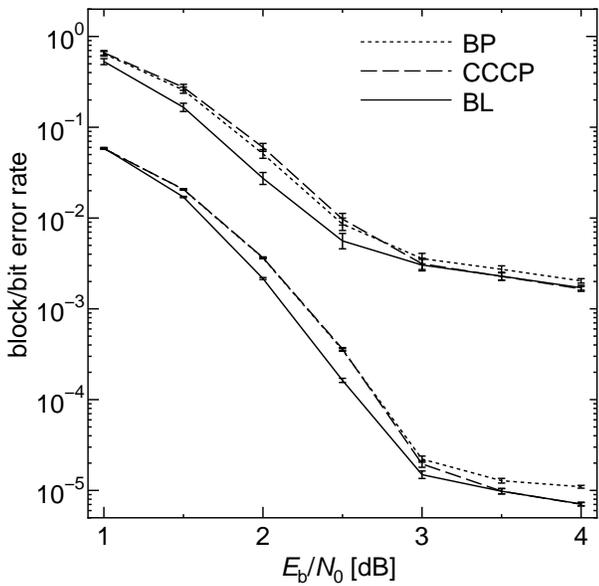}\\
    \caption{
        Block (upper three lines) / bit (lower three lines) error rates of BP, CCCP, and BL algorithms. 
        Configurations were as follows: code length $N=486$, number of parities $M=243$,
        degree of parity $|{\bm \mu}|=6$, degree of bit $|{\bm l}|=3$ ((3,6)-regular LDPCC),
        limit of outer loops: 10000 (the limit of loops in the case of BP),
        number of inner loops fixed as $6$ for CCCP and BL, and step width $k=0.3$ in BL.
        $\EbN0$[dB] is defined as $10 \log_{10} (1/(2\sigma^2 (N\!-\! M)/N))$.
        The number of communications were $10^3, 3\times10^3, 10^4, 3\times10^4, 10^5, 3\times10^5,$ and $10^6$
        for $\EbN0 = 1.0, 1.5,..., 4.0$, respectively.
        Each error bar denotes a $99\%$ confidence interval based on a binomial distribution.
    }
    \label{fig:ErrorRates}
\end{figure}
%    \vspace{5mm}
\begin{figure}[t]
    \includegraphics[width=80mm]{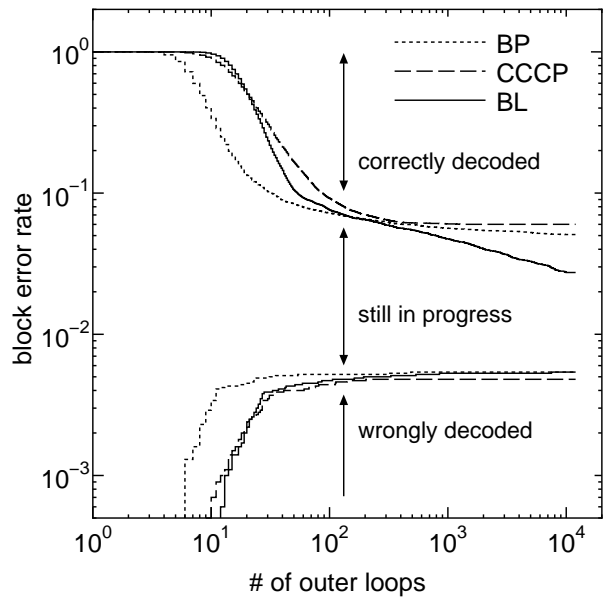}
    \caption{
        Time evolution (outer loop steps) of the not-correctly-decoded rates (upper three lines) and wrongly-decoded rates (lower three lines) for BP, CCCP, and BL algorithms.
        The data corresponds to the case of $\EbN0=2$ in Fig. \ref{fig:ErrorRates}.
    }
    \label{fig:OuterLoop}
\end{figure}
%================

  We validated the performance of the BL algorithm by comparing it with that of BP and CCCP
through a simulation of the Gallager ensemble of the short-length regular LDPCC.
As the decoding performance greatly depends on the parity check matrix,
the simulation was done over an LDPCC ensemble;
that is, we remade the matrix for every communication according to the Gallager's construction \cite{rf:Gallager}.
We assumed that the decoder knows the true noise variance of the AWGN channel, $\sigma^2$.
In the simulation, BL performed somewhat better than both BP and CCCP.

  Figure \ref{fig:ErrorRates} shows the block and bit error rates of each algorithm over various signal-to-noise ratios ($\EbN0$).
BL performed better than BP and CCCP, especially in the area where $\EbN0$ was around $2$ dB.
The error floor appeared in the area where $\EbN0$ was greater than about $2.5$ dB.
This error floor probably occurred due to the short loop of the parity check matrix.

  Figure \ref{fig:OuterLoop} shows the time (outer-loop step) evolution of the rate for both the not-correctly-decoded and wrongly-decoded cases.
In the early outer-loop steps, BP tended to reach the correct codeword first, and then CCCP and BL followed.
In the later steps, BL continuously improved the block error rate,
while CCCP had little effect after about the $500$-th step.
The effect of BP was intermediate.
The rates for the wrongly-decoded case were almost the same among the three algorithms.
These results suggest that the BL algorithm will outperform BP and CCCP if we can afford a high calculation cost -- for example, $1000$ outer-loop steps.

  The calculation cost is roughly proportional to the number of inner loops (we regard that of BP to be $1$).
So, if we set the number of inner loops as $6$ for CCCP and BL,
the cost ratio of BP, CCCP, and BL will be about $1:6:6$.
If we consider the average number of outer loops, the difference could become larger (e.g., $1:10:12$),
but this depends on the upper limit on the number of outer loops.

  Parallelization is also an important factor regarding calculation cost.
Briefly, parallelization of the BP loop is possible.
It is also possible for the outer loops of CCCP and BL, but not for the inner loop of CCCP.
On the other hand, it is indispensable for the inner loop of BL to achieve fast convergence.

  Parameter optimization of the three algorithms is a real problem.
In the case of BP, we have to determine only the upper limit of the outer loops.
For CCCP, we also have to determine the number of inner loops.
For BL, in addition to the CCCP parameters, we have to determine the step width of the natural gradient descent.
Empirically, the configuration of the step width appears rather robust since the
simulated BL performance generally exceeded that of the other algorithms (Fig. \ref{fig:ErrorRates})
even though they shared a common step width configuration (i.e., $k=0.3$).

  The optimization of the parity check matrix is also a problem, especially for short-length LDPCC.
We will further investigate the dependence of these algorithms on the matrix in our future work.

%================================================================
\section{Conclusion}
  The method we have proposed minimizes the Bethe free energy based on the Bowman-Levin (BL) approximation.
The BL algorithm combined with the natural gradient descent method successfully converges.
We have compared our BL algorithm to the belief propagation (BP) and concave convex procedure (CCCP) algorithms
with respect to the decoding problem of the Gallager ensemble of short-length regular low-density parity check codes (LDPCC)
over an additive white Gaussian noise (AWGN) channel.
Simulation showed that the BL algorithm outperformed the BP and CCCP algorithms, although the BL calculation cost was greater.
This suggests that the BL algorithm can be successfully applied to other problems to which BP or CCCP have already been applied.

%================================================================
\section*{Acknowledgements}
  This work was partially supported by a Grant-in-Aid for Scientific Research on Priority Areas No. 14084206.

%================================================================

\end{document}